\documentclass[showpacs,preprintnumbers,amsmath,amssymb,prd]{revtex4}

\usepackage{graphicx}
\usepackage{dcolumn}
\usepackage{bm}

\newcommand{\ihmpc}{\, h\, {\rm Mpc}^{-1}}

\newcommand{\lyaf}{Ly$\alpha$ forest}

\newcommand{\vp}{\mathbf{p}}
\newcommand{\vx}{\mathbf{x}}
\newcommand{\vk}{\mathbf{k}}
\newcommand{\vq}{\mathbf{q}}
\newcommand{\vnabla}{\mathbf{\nabla}}
\newcommand{\vv}{\mathbf{v}}
\newcommand{\orderfour}{\mathcal{O}\left(\delta_1^4\right)}

\begin{document}

\title{Dark matter clustering:  a simple renormalization group approach}

\author{Patrick McDonald}
\email{pmcdonal@cita.utoronto.ca}
\affiliation{Canadian Institute for Theoretical Astrophysics, University of
Toronto, Toronto, ON M5S 3H8, Canada}

\date{\today}

\begin{abstract}

I compute a renormalization group (RG) improvement to the standard 
beyond-linear-order Eulerian perturbation theory (PT) calculation of the power 
spectrum of large-scale density fluctuations in the Universe.  At $z=0$, for a 
power spectrum matching current observations, lowest order RGPT appears to be 
as accurate as one can test using existing numerical simulation-calibrated 
fitting formulas out to at least $k\simeq 0.3\ihmpc$; although 
inaccuracy is
guaranteed at some level by approximations in the calculation (which can be 
improved in the future).
In contrast, standard PT breaks down virtually as soon as beyond-linear 
corrections become non-negligible, on scales even larger than $k=0.1\ihmpc$. 
This extension in range of validity could substantially enhance the usefulness 
of PT for interpreting baryonic acoustic oscillation surveys aimed at probing 
dark energy, for example.  I show that the predicted power spectrum converges 
at high $k$ to a power law with index given by the fixed-point solution of the 
RG equation.  I discuss many possible future directions for this line of work.
The basic calculation of this paper should be easily understandable without any 
prior knowledge of RG methods, while a rich background of mathematical physics 
literature exists for the interested reader.

\end{abstract}


\maketitle

\section{Introduction}

Statistics of the large-scale density fluctuations in the Universe provide
an invaluable source of information on the very early Universe, where the 
initial density perturbations that grow by gravitational instability were 
created, and on the material content and gravitational laws of the Universe, 
which determine the growth of structure and how it looks to us from a distance
\citep{2006JCAP...10..014S}.
The standard practical toolbox we use for interpreting observations consists
of linear theory on very large scales or at very early times
\citep{1996ApJ...469..437S}, and everywhere 
else numerical simulations which compute the fully non-linear evolution of a 
realization of the density field in a patch of Universe.
Fitting formulas, motivated by physical intuition but ultimately heuristic,
are often used to more efficiently interpolate between simulation 
results \citep{2003MNRAS.341.1311S,2006ApJ...646..881W}.  

Beyond linear order perturbation theory (PT) has been considered for a long 
time \citep{1980lssu.book.....P,1981MNRAS.197..931J,1983MNRAS.203..345V,
1984ApJ...279..499F,
1986ApJ...311....6G,1994ApJ...431..495J,1996ApJ...473..620S}, but does not 
currently play an important role in our
interpretation of precision observations
(see \cite{2002PhR...367....1B} for a thorough review of large-scale structure
perturbation theory).
The desire to measure dark energy properties using observations of baryonic 
acoustic oscillation features \citep{1998ApJ...504L..57E,
1998ApJ...496..605E,2001ApJ...557L...7C,2003astro.ph..1623E,
2003ApJ...594..665B,2003PhRvD..68h3504L,2003ApJ...598..720S,
2004ApJ...615..573M,2005ApJ...631....1G,2005astro.ph..7457G,
2005MNRAS.357..429A,2005MNRAS.363.1329B,2006MNRAS.365..255B,
2006MNRAS.366..884D,2006astro.ph..7122M}, which fall more or
less exactly in the range of scales where higher order perturbation
theory could be most useful, gives this line of work much more pressing 
practical
relevance than it has had in the past \citep{2006ApJ...651..619J}.  
More traditional uses of galaxy clustering to measure cosmological parameters,
long reliant on the assumption that galaxy power 
is simply proportional to linear mass power, have also reached the point where 
weakly non-linear effects are a substantial limitation 
\citep{2004ApJ...606..702T,2006MNRAS.366..189S}.
Additionally, other probes of large-scale structure, like the \lyaf\ 
\citep{2005ApJ...635..761M,2006MNRAS.365..231V},
weak lensing \citep{2006ApJ...647..116H}, 
galaxy cluster/Sunyaev-Zel'dovich effect (SZ) measurements 
\citep{2005astro.ph.11060D}, and possibly 
future 21cm surveys \citep{2005MNRAS.364..743N}, could all benefit from 
improved computational techniques. 

The problem with higher order perturbation theory is that it does not work 
very well in exactly the regime where it could currently be most useful:
at low redshift and wavenumber $k\sim 0.1\ihmpc$, where galaxy clustering
is significantly but still only weakly non-linear \citep{2004PhRvD..70h3007S}.
Recently, \cite{2006PhRvD..73f3520C,2006PhRvD..73f3519C} proposed a
renormalization technique aimed at improving PT calculations.  They re-sum an 
infinite series of perturbation theory
terms (Feynman diagrams) which determine the memory of perturbations to initial
conditions as a function of scale (see also \cite{2001NYASA.927...13S}).
This method is apparently 
computationally difficult because they have not yet computed a power spectrum
using it.  The renormalization group (RG) method I present here is different, 
but surely related.
There has been some work toward using RG techniques to 
model large-scale structure \citep{1997EL.....38..637B,1999A&A...344...27D,
2001IJMPA..16.2041G}.  
The methods used are different than mine, and 
have not yet seen much practical application for comparison with 
observations.

My implementation of the RG is inspired more by its use for dealing
with secular divergences in perturbative solutions of differential equations 
\citep{1994PhRvL..73.1311C,1996PhRvE..54..376C} than by the probably better
known applications 
in high energy and statistical physics \citep{1974PhR....12...75W,
1983RvMP...55..583W,1994RvMP...66..129S,2002PhR...363..223B}, although all of
these uses are related.
The basic idea is to identify a quantity in the problem that is not directly 
observable and modify it to remove breakdowns in the perturbation 
theory, leading to a well-behaved renormalized observable. 
In this paper I renormalize the initial (linear theory) power spectrum to 
remove the relative (not actually infinite for realistic power spectra)
divergence of the lowest order correction to linear theory.  
More detailed mathematical clarification and explanation
of the meaning of the method of 
\cite{1994PhRvL..73.1311C,1996PhRvE..54..376C}, in terms of the theory of 
envelopes, can be found in 
\cite{1995PThPh..94..503K,1997PThPh..97..179K}.
Even more detailed treatments can be found in \citep{2001PhR...352..219S,
2006JPhA...39.8061K}.
These papers should be useful to the
reader who is unsatisfied with my presentation, or wants to extend it.  
During the last several years,  
\cite{1999PhRvD..60f5003B,1999PhRvD..59j5019B,2000PhRvD..61f5006B,
2000PhRvD..62j5026W,2002PhRvD..65d5014B,2003PhRvD..67f5022B,
2003AnPhy.307..335B} have extended and
generalized this dynamical renormalization group method to 
compute the real time evolution of expectation values of quantum fields. 

The rest of the paper is as follows:
In \S \ref{seccalc} I derive a simple formula for the RG-improved power 
spectrum of mass density fluctuations.  In \S \ref{secresults} I show 
numerical results and compare them to standard PT and fitting formulas 
calibrated using simulations.
Finally, in \S \ref{secdiscuss} I discuss possible future directions for this
kind of work.

\section{Calculation \label{seccalc}}

In this section I present the basic calculation.  
In \S \ref{secevolutioneqs} I discuss the time evolution equations for cold 
dark matter, which are the starting point for perturbation theory.
In \S \ref{secstandardpert}
I review standard Eulerian perturbation theory.  
Finally, in \S \ref{secRG} I
apply the renormalization group improvement.

\subsection{Evolution equations \label{secevolutioneqs}}

We are interested in the statistics of the mass density field, 
$\delta(\vx,\tau)=\rho(\vx,\tau)/\bar{\rho}-1$, with Fourier transform 
$\delta(\vk,\tau)=\int d^3\vx~\exp(i \vk\cdot \vx)~\delta(\vx,\tau)$, 
where $\vx$ is the comoving position and 
$\tau=\int dt/a$ is the conformal time, with $a=1/(1+z)$ the expansion factor.
In particular, I will compute the power spectrum, $P(k,\tau)$, defined by 
\begin{equation}
\left<\delta(\vk,\tau) \delta(\vk',\tau)\right> = 
(2\pi)^3 \delta_D(\vk+\vk')P(k,\tau) 
\end{equation}

For now I will assume only cold dark matter, i.e.,  collisionless
particles with insignificant initial velocity dispersion.  The exact 
evolution is is described by the Vlasov equation \citep{1980lssu.book.....P}
\begin{equation}
\frac{\partial f}{\partial \tau} +\frac{1}{a~m}~\vp\cdot\vnabla f
-a~m~\vnabla \phi \cdot \vnabla_\vp f=0 ~,
\label{eqvlasov}
\end{equation}
with 
\begin{equation}
\nabla^2\phi = 4~\pi~G~a^2 \bar{\rho}~\delta~,
\label{potential}
\end{equation}
where $f(\vx,\vp,\tau)$ is the particle density at phase-space position 
$(\vx,\vp)$,
$m$ is the particle mass (which plays no role in the final results), and 
$\vp= a~m\vv$ ($\vv$ here is a particle's peculiar velocity, not to be
confused with the mean peculiar velocity of all particles at some point in
space, used everywhere else). 
Except when otherwise indicated $\vnabla=\mathbf{\partial/\partial x}$.
The density field is obtained by averaging the distribution function over 
momentum:
\begin{equation}
\rho(\vx,\tau)=m a^{-3}\int d^3 p~f(\vx,\vp,\tau)~,
\end{equation}    
and the bulk (mean) velocity and higher moments of the velocity distribution
e.g., the dispersion of particle velocities around their bulk velocity, can be 
similarly obtained by multiplying the distribution function by any number of
$\vp$'s (e.g., one to obtain bulk velocity) before integrating over $\vp$.  
As discussed by \cite{1980lssu.book.....P}, taking moments of 
the Vlasov equation with respect to momentum leads to a hierarchy of 
evolution equations for these quantities.
Dropping the higher moments of the velocity distribution gives the usual 
hydrodynamic equations for density and bulk velocity: 
\begin{equation}
\frac{\partial \delta}{\partial \tau}+
\vnabla\cdot\left[\left(1+\delta\right)\vv\right]=0~,
\label{eqcontinuity}
\end{equation}
and
\begin{equation}
\frac{\partial \vv}{\partial \tau}+\mathcal{H} 
\vv+\left(\vv\cdot \vnabla\right)\vv=-\vnabla \phi ~,
\label{eqeuler}
\end{equation}
where $\mathcal{H}=d\ln a/d\tau=a H$ with $H$ the usual Hubble parameter.

\subsection{Standard Eulerian perturbation theory \label{secstandardpert}}

Perturbation theory consists of writing the density and velocity fields as
a series of terms of at least formally increasing order of smallness, i.e., 
$\delta=\delta_1+\delta_2+\delta_3+...$.  The evolution equations are solved
order-by-order, with lower order solutions appearing as sources in the 
higher order equations so that $\delta_n$ is of order $\delta_1^n$
\citep{2002PhR...367....1B}.
The power spectrum for Gaussian initial conditions is given by
\begin{equation}
\left<\delta(\vk) \delta(\vk')\right> = 
\left<\delta_1(\vk) \delta_1(\vk')\right> +
\left<\delta_1(\vk) \delta_3(\vk')\right> +
\left<\delta_3(\vk) \delta_1(\vk')\right> +
\left<\delta_2(\vk) \delta_2(\vk')\right> +...
\end{equation}
where no terms 3rd order in $\delta_1$ appear because the expectation value of 
any term cubic in a Gaussian field is zero.

At this point I assume an Einstein-de Sitter Universe for simplicity, and 
define
\begin{equation}
P(k,\tau)= 
D^2(\tau) P_{11}(k) + D^4(\tau) \left[P_{13}(k)+P_{22}(k)\right]+...~,
\label{Pspeceq}
\end{equation}
where $D(\tau)=\delta_1(\tau)/\delta_{\rm initial}$ is the linear theory 
growth factor.  The Einstein-de Sitter assumption is needed to avoid more 
complicated time dependence of $P_{13}$ and $P_{22}$, but
the real Universe is of course not Einstein-de Sitter.  
Fortunately, this is 
not a significant problem because, to percent level accuracy 
\citep{1991ApJ...371....1M,1992ApJ...394L...5B,1994ApJ...433....1B,
1995A&A...296..575B,1995PThPh..94.1151M,1995MNRAS.276...39C,
1998ApJ...496..586S,2006ApJ...651..619J}, 
the effect of changing the background
model can be included by simply using the correct linear growth factor in 
Eq.~(\ref{Pspeceq}).
I will sometimes refer to $P_{13}$ and $P_{22}$ as 2nd order, meaning in the
initial power spectrum amplitude, not to be confused with the fact that they are
4th order in $\delta_1$ and require calculating the evolution of $\delta$ to
3rd order, i.e., $\delta_3$.
\cite{1992PhRvD..46..585M} derived the following useful form of the equations 
for $P_{13}(k)$ and $P_{22}(k)$:
\begin{equation}
P_{13}(k)=\frac{k^3 P_{11}(k)}{252 \left(2 \pi\right)^2}\int_0^\infty dr 
P_{11}(k r)
\left[\frac{12}{r^2}-158+100 r^2 -42 r^4 +\frac{3}{r^3}\left(r^2-1\right)^3
(7 r^2 +2)\ln\left|\frac{1+r}{1-r}\right|\right]
\label{P13}
\end{equation}
and 
\begin{equation}
P_{22}(k)=\frac{k^3}{98 \left(2 \pi\right)^2}\int_0^\infty dr P_{11}
\left(k r\right)
\int_{-1}^1 dx P_{11}\left[k\left(1+r^2-2 r x\right)^{1/2}\right]
\frac{\left(3 r +7 x-10 r x^2\right)^2}{\left(1+r^2-2 r x\right)^2}~.
\label{P22}
\end{equation}
Note that, due to many cancellations, these terms are not as divergent as they
might appear at first glance, their sum being convergent at high $k$ for 
power law $P_{11}(k)$ with $n=d\ln P/d\ln k<-1$ 
and at low $k$ when $n>-3$ \citep{1992PhRvD..46..585M}.
Their sum is zero for $n\simeq-1.4$ \citep{2002PhR...367....1B}, a fact that
will have interesting consequences for our RG calculation.

\subsection{Renormalization group improvement \label{secRG}}

The problem with the standard calculation outlined in \S\ref{secstandardpert}, 
which leads us to renormalization,
is that the 2nd term in Eq.~(\ref{Pspeceq}) diverges relative to the 
first (although it does not literally become infinite for realistic power
spectra), at increasingly large scales (small $k$) as time progresses.
This divergence is not exactly unphysical -- we expect 
non-linearities to become important during gravitational collapse -- however,
the basic premise of perturbation theory is violated when higher order terms 
become large.  There is no {\it a priori} reason to expect the results to be
accurate.
I employ a renormalization group calculation to cure this divergence.
I start in \S\ref{simpRG} with a quick derivation motivated by the method of 
\cite{1994PhRvL..73.1311C}.  Then I discuss the mathematical background, 
physical interpretation, approximations, and limitations of the method in 
\S\ref{secexplanation}.       

\subsubsection{Simple calculation \label{simpRG}}

To simplify the presentation of the calculation,
I rewrite Eq.~(\ref{Pspeceq}) in a
more compact, schematic form,
\begin{equation}
\tilde{P}(k,\tau)\simeq P_L(k) + A~\left[P_L^2\right](k)~,
\end{equation}
where $A\propto D^2(\tau)$, $\tilde{P}(k,\tau)\equiv P(k,\tau)/A$,
$P_L(k)$ is the initial condition power,
and $[P_L^2](k)\equiv P_{13}(k)+P_{22}(k)$ is the higher order correction term, 
which is 
quadratic in $P_L(k)$, as given by Eqs.~(\ref{P13}) and (\ref{P22}). 
I now follow the method described in \cite{1994PhRvL..73.1311C} to 
deal with the problem of the 2nd order term  becoming large (deferring a
detailed explanation of the meaning of this method to \S\ref{secexplanation}).
I rewrite $A=A-A_\star+A_\star$, where 
$A_\star$ is just an arbitrary constant, so that
\begin{equation}
\tilde{P}(k,\tau)=P_L(k)+A_\star \left[P_L^2\right](k)+ 
(A-A_\star)\left[P_L^2\right](k)~.
\end{equation}  
I then absorb the constant $A_\star \left[P_L^2\right](k)$ 
into the initial
conditions, $P_L(k)$, defining the renormalized initial conditions
$P_\star(k,A_\star)$, so that
\begin{equation}
\tilde{P}(k)=P_\star(k,A_\star)+(A-A_\star) \left[P_\star^2\right](k)~,
\label{eqrenormed}
\end{equation} 
where I can freely replace $P_L$ with $P_\star$ in the 2nd term because
the change is formally 3rd order (in the initial power spectrum amplitude). 
The observable power spectrum $\tilde{P}(k)$ obviously must be
independent of the completely artificial constant $A_\star$, so I impose
$d\tilde{P}(k)/dA_\star=0$ and find
\begin{equation}
\frac{dP_\star(k,A_\star)}{dA_\star}=\left[P_\star^2\right](k)~,
\label{eqbeta}
\end{equation}
where I have dropped the derivative of $\left[P_\star^2\right](k)$ because it
is higher order (this is seen to be self-consistent from Eq.~\ref{eqbeta}
itself).  
This equation is easy enough to solve numerically, given  
$P_\star(k,A_\star)$ at some value of $A_\star$.
Since $A_\star$ is completely arbitrary, I will in the end choose
$A_\star=A$, removing the 2nd term in Eq.~(\ref{eqrenormed}) entirely.  
The result for $\tilde{P}(k,\tau)$ is then simply 
$P_\star(k,A_\star=A)$, obtained by solving Eq.~(\ref{eqbeta}).
To reproduce the linear theory result at very early times,
the initial conditions for the solution of Eq.~(\ref{eqbeta}) must be
$P_\star(k,A_\star=0)=P_L(k)$.

\subsubsection{Explanation \label{secexplanation}}

The method of \cite{1994PhRvL..73.1311C} and related papers was, by the authors
own admission, somewhat lacking in rigorous motivation.  This situation was 
clarified by others \cite{1995PThPh..94..503K,1997PThPh..97..179K,
2001PhR...352..219S,2006JPhA...39.8061K}; however, 
their explanations may be opaque.  A very simple example should serve to 
explain and clarify the method.  Consider the differential equation
\begin{equation}
\dot{\delta}=\delta+\delta^2.
\label{eqverysimple}
\end{equation}
The exact solution is easy to obtain, but suppose one did not know that and
solved Eq. (\ref{eqverysimple}) using perturbation theory for small initial
$\delta$, as in the cosmological situation.
The solution to the linearized equation is $\delta_1 = g_1 e^t$.  
The equation for the first non-linear correction, $\delta_2$, is then
\begin{equation}
\dot{\delta}_2=\delta_2 +g_1^2 e^{2 t}
\end{equation}
with solution 
\begin{equation}
\delta_2=g_2 e^t +g_1^2 e^{2 t}.
\end{equation}
The approach generally followed in the cosmological case is to set
the term that is the solution to the homogeneous equation, $g_2 e^t$, to zero, 
i.e., to assume
that at asymptotically early times $\delta\rightarrow\delta_1$; however, this 
is a {\it choice}, not necessary.  The key observation here is that solving two 
differential equations has 
produced two parameters in the solution, $g_1$ and $g_2$, but we only 
need one to satisfy the 
physical boundary conditions.  The RG method exploits and fixes the otherwise 
arbitrary extra parameter of the perturbative solution (note that
this approach gives exactly the same result as
the approach of ``splitting'' the linear solution into two
pieces, one of which is considered to be second order and used to cancel the 
non-linear correction).

For the standard choice $g_2=0$, 
$\delta_2$ inevitably becomes larger than $\delta_1$, invalidating the 
perturbation theory.  Note, however, that using $g_2$ we can actually guarantee
that at any one chosen time, $t_\star$, $\delta_2(t_\star)=0$, and very near 
$t_\star$, $\delta_2 \ll \delta_1$.  
This requires $g_2=-g_1^2 e^{t_\star}$, which 
leads to full solution
\begin{equation}
\delta\simeq g_1 e^t + g_1^2 e^t(e^t-e^{t_\star}).
\label{eqsimple2ndordersolution}
\end{equation} 
Clearly the value of $g_1$ must depend on the choice of $t_\star$, i.e., 
$g_1=g_1(t_\star)$, 
for a fixed physical situation (fixed full solution $\delta(t)$); however,
nothing would be accomplished by simply fixing $g_1(t_\star)$ to match the 
initial
conditions, because this generally requires using the perturbative solution 
outside its range of validity. 
The RG method is to determine $g_1(t_\star)$ by enforcing the idea that the 
physical solution should be 
independent of the arbitrary time $t_\star$, while using the perturbative 
solution only for $t$ infinitesimally close to $t_\star$, where it should be 
valid, i.e., \begin{equation}
\left. \frac{d\delta(t)}{d t_\star}\right|_{t_{\star}=t}=0=
\frac{d g_1(t)}{d t} e^t-g_1^2 e^{2 t} 
\end{equation}
or
\begin{equation}
\frac{d g_1(t)}{d t} =g_1^2 e^{t}
\label{eqverysimplebeta}
\end{equation} 
The exact solution to this equation is $g_1(t)=c/(1-c e^{t})$ 
where $c$ is a 
constant which will allow matching of the physical initial conditions. 
Finally, referring back to Eq. (\ref{eqsimple2ndordersolution}), evaluated
at $t_\star=t$, gives the final solution
\begin{equation}
\delta(t)= \frac{c e^t}{1-c e^t}.
\end{equation}
While such good fortune would generally not be expected, it happens that this 
is the 
exact solution to the original differential equation, Eq. (\ref{eqverysimple}).

The intuitive picture one should have of this calculation is of stepping 
forward
in time slightly using the perturbative solution centered at the present time,
then taking the result and using it as the initial condition for a perturbative
solution around the new time, followed by another step, and so on. 
\cite{1995PThPh..94..503K} showed that this method gives the
envelope equation for the set of solutions obtained by applying the 
perturbation theory near various times $t_\star$ (the envelope
is the curve which is tangent to each of the different solutions).

To make the analogy with original case of the power spectrum clear, note that 
changing 
variables to $A=\exp\left(t\right)$, $A_\star=\exp\left(t_\star\right)$, and
$\tilde{\delta}=\delta/A$ in Eq. (\ref{eqsimple2ndordersolution}) produces
an equation with the same form as Eq. (\ref{eqrenormed}).  
The rest of the derivation follows identically until, unfortunately, 
Eq. (\ref{eqbeta}) is not analytically solvable because $P(k)$ represents an
infinite number of variables rather than one. 

While the resulting equations look similar, the skeptical reader may still 
question the relation between the derivation of Eq.~(\ref{eqbeta}) for the
power spectrum and Eq. (\ref{eqverysimplebeta}) for a single variable governed
by a differential equation, e.g., $\delta$ would be most naturally be 
identified with the cosmological density field, not the power spectrum. 
To complete the connection one must return to the original derivation of 
cosmological perturbation theory.  At linear order $\delta_1(\vk)=D~ g_1(\vk)$ 
where 
$g_1(\vk)$ is the normalization and $D=a$ is the growth factor (assuming 
an EdS Universe as usual), and I have dropped
the decaying mode solution as usual (I will discuss this further  
below).  At 2nd order 
\begin{equation}
\delta_2(\vk)= D~g_2(\vk)+
D^2 \int \frac{d^3 \vq}{\left(2 \pi\right)^3}g_1(\vq) g_1(\vk-\vq)
J_S^{(2)}(\vq,\vk-\vq)~,
\end{equation}
where the $D~g_2(\vk)$ term is set to zero in the standard calculation
(see \cite{1998MNRAS.301..797H} for explicit forms of 
$J_S^{(2)}$ and $J_S^{(3)}$ below).  Again,
this equation is missing terms which grow like $D^{-1/2}$ or slower
(naively one would expect them to be unimportant).

$\delta_3$ is needed to compute the power spectrum, but computing the 
bispectrum at this point is informative:
\begin{eqnarray}
\left<\delta(\vk_1)\delta(\vk_2)\delta(\vk_3)\right>&=&
D^3\left<g_1(\vk_1)g_1(\vk_2)g_1(\vk_3)\right>+
D^3\left<g_2(\vk_1)g_1(\vk_2)g_1(\vk_3)\right> \nonumber \\&+&
D^4 \int \frac{d^3 \vq}{\left(2 \pi\right)^3}
J_S^{(2)}(\vq,\vk_1-\vq) \left<g_1(\vq) g_1(\vk_1-\vq) g_1(\vk_2) g_1(\vk_3)
\right>
+{\rm cyclic~ permutations~ of}~\vk_i~.
\label{eqbispectrum}
\end{eqnarray}
The last term here is the usual perturbative result.
The first term is normally zero, for Gaussian initial conditions.  The middle 
term is normally zero because $g_2(\vk)$ is chosen to be zero.
One can look at Eq. (\ref{eqbispectrum}) two ways:  
The easy way is to say that there is no breakdown in perturbation theory, where
a higher order term becomes larger than a lower order term.  The last term is
the leading term, i.e., the bispectrum is fundamentally $\orderfour$, and there 
is no need for the first two terms to enter this discussion at all.
The second approach is to argue
that all possible terms should be included in the calculation -- the
fact that the first term happens to be zero initially is a peculiarity of the
initial conditions, not fundamental to the calculation.  In this  
approach, the new
middle term, since it is arbitrary, could be set in a way that would 
cancel the last term at some chosen $D_\star$.  
The first term becomes a function of $D_\star$, and requiring that the 
solution be independent of $D_\star$ would produce an RG equation for 
$\left<g_1(\vk_1)g_1(\vk_2)g_1(\vk_3)\right>(D_\star)$,
just as above in the simple example.  
The second approach is clearly more
thorough, however, in this paper I will take the easy route and declare the 
usual bispectrum term to be the leading order, with no need for 
renormalization, i.e., I will set $g_2(\vk)\equiv 0$ as usual (we will see that
this term is not needed to renormalize the power spectrum).
While my approach may not give the best possible result, it is not
fundamentally incorrect, at least not any more than standard perturbation 
theory is fundamentally incorrect.  Renormalization is a tool, not an 
imperative, and renormalizing part of the calculation but not all of it 
should be better than nothing.  

Moving on to $\delta_3$, and again including the homogeneous solution, gives
\begin{equation}
\delta_3(\vk)=D g_3(\vk)+D^3
\int \frac{d^3 \vq_1}{\left(2 \pi\right)^3}
\frac{d^3 \vq_2}{\left(2 \pi\right)^3}
g_1(\vq_1) g_1(\vq_2) g_1(\vk-\vq_1-\vq_2)
J_S^{(3)}(\vq_1,\vq_2,\vk-\vq_1-\vq_2)~,
\label{eqdelta3}
\end{equation}
where terms that grow like $D^{1/2}$ or slower have as usual been dropped.
The power spectrum is now
\begin{equation}
\left<\delta(\vk_1)\delta(\vk_2)\right>=
D^2 \left<g_1(\vk_1)g_1(\vk_2)\right>+
D^2 \left<g_3(\vk_1)g_1(\vk_2)+
g_1(\vk_1)g_3(\vk_2)\right>+
D^4 (2 \pi)^3 \delta^D(\vk_1+\vk_2) [g_1^4](k_1)
\end{equation}
where $[g_1^4](k)\equiv P_{13}(k)+P_{22}(k)$ is, 
as $[P_L^2](k)$ above, just a quick way of writing
the usual $\orderfour$ perturbative correction terms (note that if I had 
promoted the leading order bispectrum to $\mathcal{O}(\delta_1^3)$ there would 
be more new terms in the power spectrum).
Following the simple example above, we know just what to do here:  eliminate 
the 
$\orderfour$ term at $D_\star$ by a specific choice of $g_3(\vk)$, i.e.,
\begin{equation}
\left<g_3(\vk_1)g_1(\vk_2)+
g_1(\vk_1)g_3(\vk_2)\right>=-
D_\star^2 (2 \pi)^3 \delta^D(\vk_1+\vk_2) [g_1^4](k_1)~,
\end{equation}
which finally leads to
\begin{equation}
P(k)=D^2 P_{11}(k)+D^2(D^2-D^2_\star)[P_{13}(k)+P_{22}(k)]~,
\end{equation}
where $P_{11}(k)$ must be a function of $D_\star$.
After the same renamings, this is exactly the starting point for 
the original RG calculation above, Eq. (\ref{eqrenormed}).

While the approach described here apparently takes care of the relative 
divergence of the corrections to the
power spectrum of density fluctuations, there is a hidden approximation that 
guarantees some imperfection.  In the usual calculation the initial conditions
are specified by simply specifying the density-density power spectrum 
$P^{\delta \delta}(k)$, because for growing mode initial conditions the 
power spectrum of velocity divergence fluctuations 
$\theta\equiv-\mathcal{H}^{-1}\vnabla\cdot\vv$, 
$P^{\theta \theta}(k)$, and the cross-correlation between $\delta$ and 
$\theta$, $P^{\delta \theta}$, are equal to $P^{\delta \delta}(k)$.
However, the non-linear corrections to these three spectra are not equal, i.e.,
decaying modes are created.  To simultaneously cancel the corrections to all 
three spectra,
decaying modes would need to be included in the solutions for $\delta$,
e.g., $\delta_1=g_1(\vk) D + d_1(\vk) D^{-3/2}$.
They do not decay instantly, so the non-linear terms they produce 
can not be safely dropped.   
It is hard to estimate how large the net effect of these changes would be 
without actually doing the full calculation, 
although \citep{2001NYASA.927...13S} showed that the decaying mode 
contribution is a modest, although not negligible, fraction of the non-linear
evolution (in a somewhat different context).
Extension of the calculation in this paper to properly account for decaying 
modes should not be too difficult.  One simply needs to write down solutions 
for $P^{\delta \delta}(k)$, $P^{\theta \theta}(k)$, and 
$P^{\delta \theta}(k)$, including the possibility of decaying modes in the 
initial conditions.  This will produce terms similar in form to $P_{13}(k)$ 
and $P_{22}(k)$, with various changes in numerical factors.  It will then be
possible to simultaneously cancel all of the corrections to 
$P^{\delta \delta}(k)$, $P^{\theta \theta}(k)$, and $P^{\delta \theta}(k)$, 
and obtain RG equations for each of them. 
If the basic principle behind this method is useful, doing the
calculation more carefully should result in even better accuracy.

While the potential usefulness of this RG method appears to be
clear, and versions of it have been successful in many previous applications 
\citep{1990PhRvL..64.1361G,1994PhRvL..73.1311C,1995PThPh..94..503K,
1996PhRvE..54..376C,
1997PThPh..97..179K,2000AnPhy.280..236E,
2001PhR...352..219S,2006JPhA...39.8061K,1999PhRvD..60f5003B,
1999PhRvD..59j5019B,2000PhRvD..61f5006B,
2000PhRvD..62j5026W,2002PhRvD..65d5014B,2003PhRvD..67f5022B,
2003AnPhy.307..335B},
the exact conditions in which it will succeed or fail are not 
completely understood.  However, the presence of an attractive fixed point 
(or manifold) is clearly a positive feature 
\cite{1995PThPh..94..503K,1997PThPh..97..179K}.
Eq. (\ref{eqbeta}) has such a fixed point at $n=-1.4$, which appears to be 
attractive (I haven't proved this but it works for every power spectrum I've
tried).  
In the asymptotic (long time) limit the arbitrarily many degrees of 
freedom in a general power spectrum are reduced to a single amplitude of this
power law.  
The system is naturally driven towards a regime where the perturbation
theory approximation works best, i.e., the 2nd order correction to the linear
theory power spectrum vanishes.  
Furthermore, the correction to the bispectrum
also vanishes for $n=-1.4$, so we are driven to a place where the Gaussianity 
of the field is well-preserved. 
On the other hand, in the simple example of
Eq. (\ref{eqverysimple}) we obtained an exact solution despite the fact that 
there is no fixed point -- in fact, the relative size of the non-linear term
diverges with time. 
Ultimately, the best way to see whether this rearrangement of the perturbation
series is useful or not is probably to simply compute higher order terms to
see if the result converges better than standard PT or not.  

\section{Results \label{secresults}}

\begin{figure}
\resizebox{\textwidth}{!}{\includegraphics{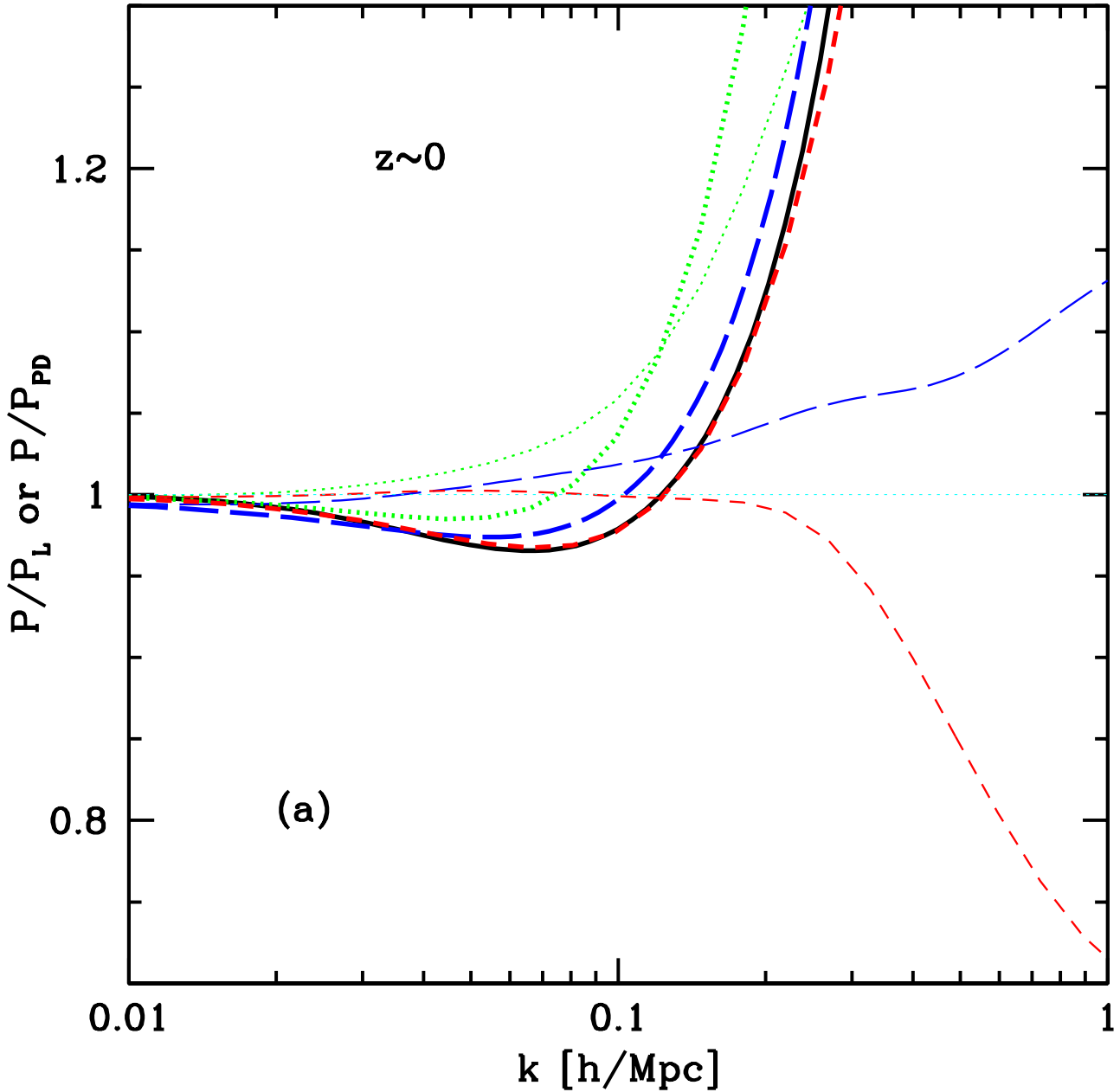}}
\caption{
Thick lines show $P(k)/P_L(k)$ at $z=0$ for the Peacock and Dodds (PD96)
fitting formula \cite{1996MNRAS.280L..19P} (black, solid), 
\cite{2003MNRAS.341.1311S}'s HALOFIT formula (blue, long-dashed), 
the RG-improved PT of this paper (red, short-dashed), and standard PT 
(green, dotted).  Thin lines show 
the other cases divided by the prediction of PD96. 
(a) and (b) are similar except for the axis scales.  Note that standard PT was
used to calibrate HALOFIT at $k<0.1\ihmpc$.
}
\label{z0}
\end{figure}

\begin{figure}
\resizebox{\textwidth}{!}{\includegraphics{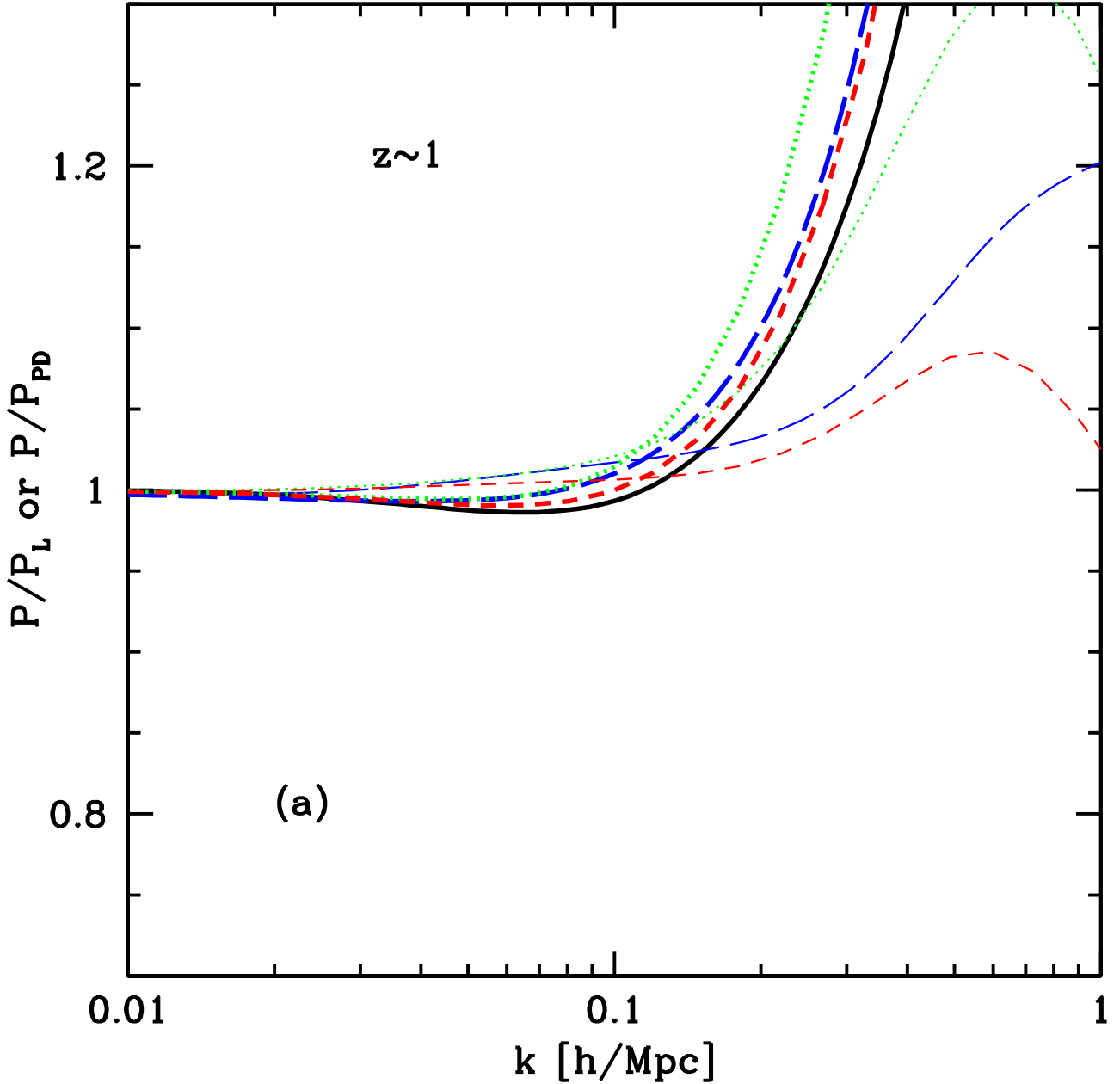}}
\caption{
As Fig. \ref{z0} except at $z\sim 1$.
We label this and subsequent $z>0$ figures by the redshift at which the
amplitude of the linear theory power in a flat $\Lambda$CDM model with
$\Omega_m=0.281$ matches the amplitude in the calculation.  The actual
redshift in the EdS model used for the calculation is somewhat lower (e.g.,
for equal $z=0$ linear power, $z=1$ in the realistic $\Lambda$CDM model 
corresponds to $z=0.61$ in the EdS model).
}
\label{z1}
\end{figure}

\begin{figure}
\resizebox{\textwidth}{!}{\includegraphics{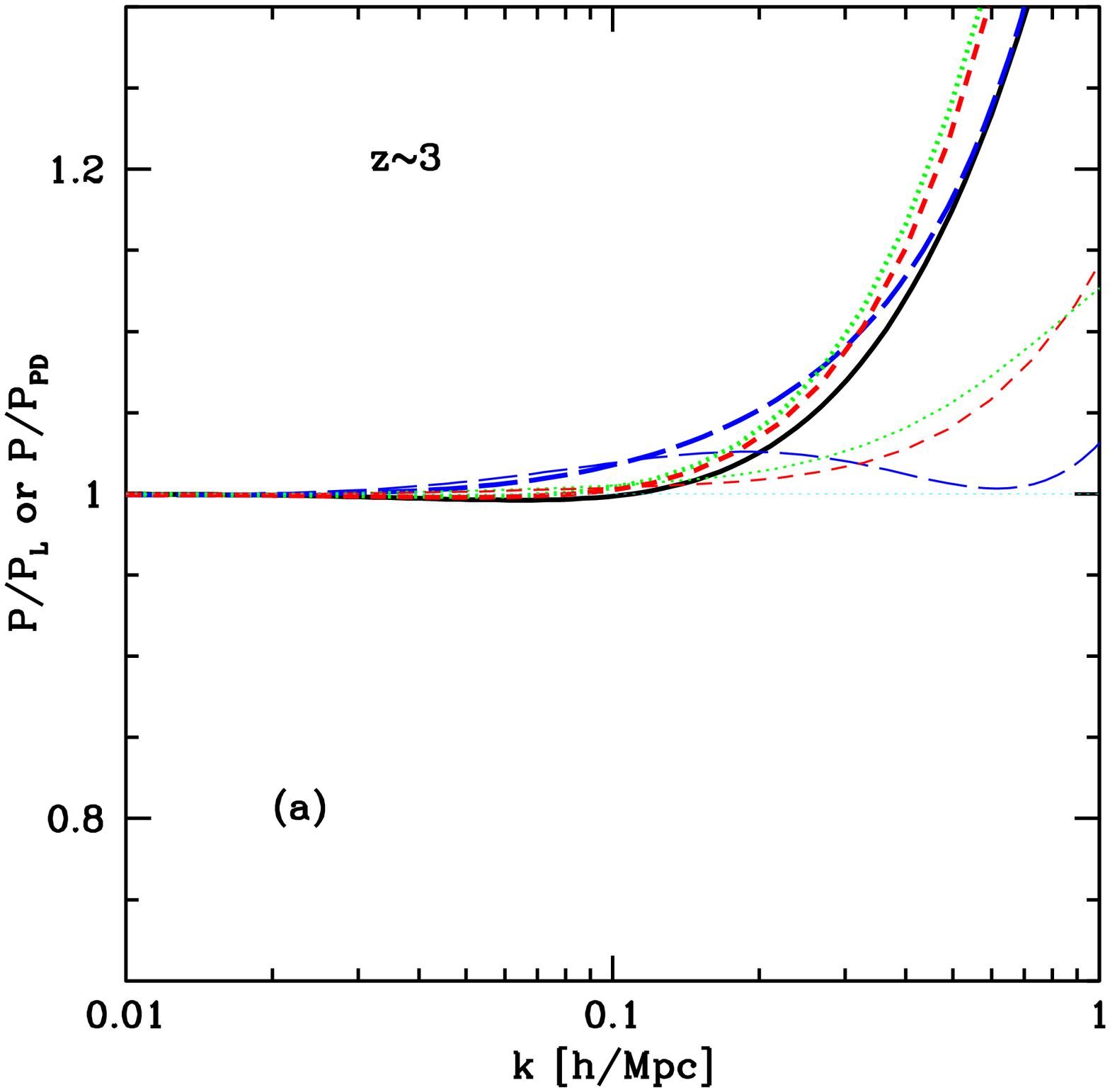}}
\caption{
As Fig. \ref{z0} except at $z\sim 3$ ($z=2.08$ in the EdS model).
}
\label{z3}
\end{figure}

\begin{figure}
\resizebox{\textwidth}{!}{\includegraphics{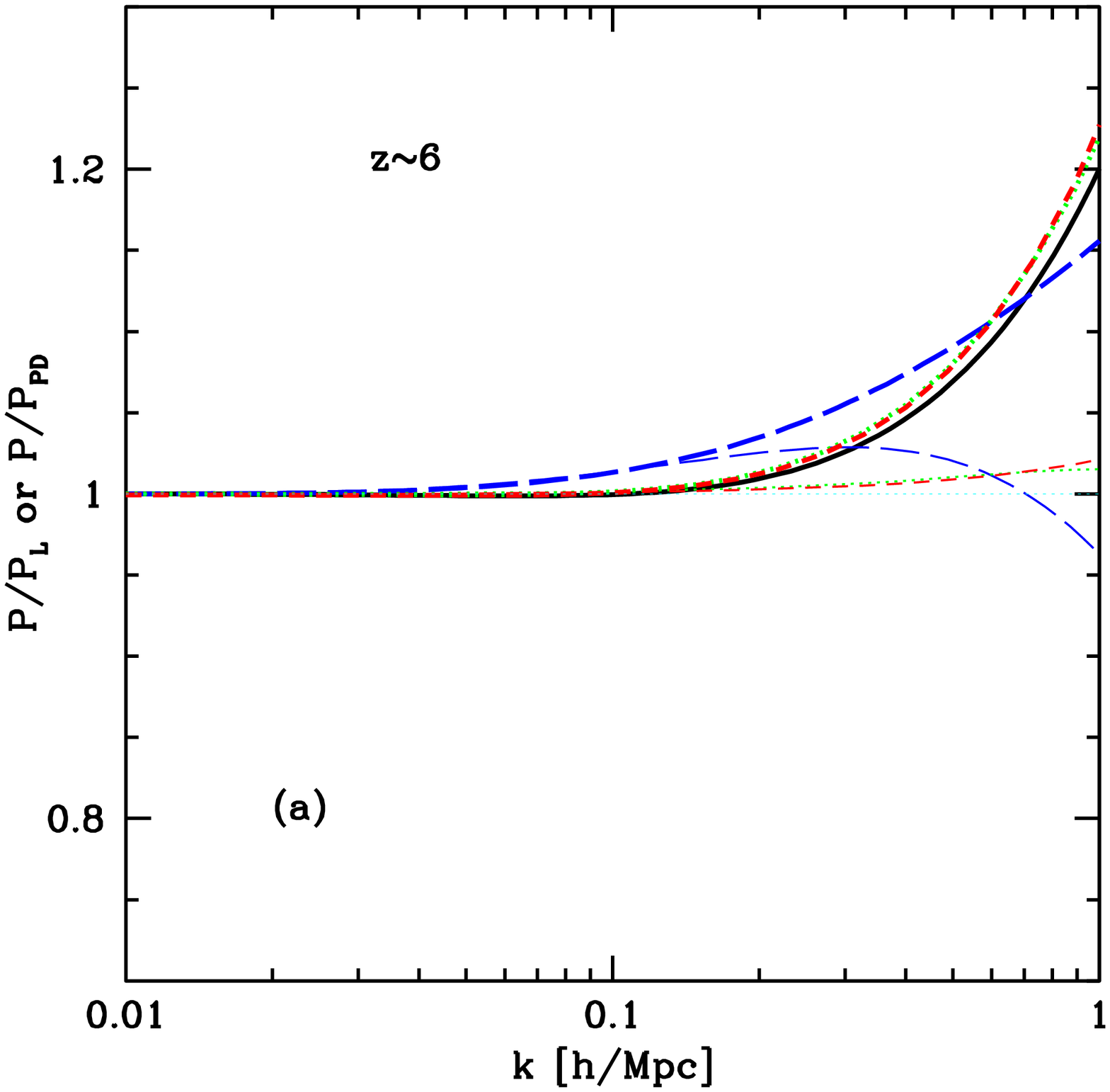}}
\caption{
As Fig. \ref{z0} except at $z\sim 6$ ($z=4.37$ in the EdS model).
}
\label{z6}
\end{figure}

In Figs. \ref{z0}-\ref{z6}
I show the results obtained by solving Eq.~(\ref{eqbeta}).
As discussed above, I am assuming
an Einstein-de Sitter Universe for the background time evolution, to avoid even
small time dependence of the $\left[P_L^2\right](k)$ term.  
I compare to the simulation-calibrated fitting formulas
of \cite{2003MNRAS.341.1311S}, including both their implementation of the 
formula from Peacock and Dodds (hereafter PD96) \cite{1996MNRAS.280L..19P} 
(note that they did not redetermine the 
parameters using their expanded set of simulations) 
and their newer HALOFIT formula.
These formulas were not calibrated for models with baryonic acoustic 
oscillations, so I compare using their standard transfer function from
\cite{1984ApJ...285L..45B}.
I choose parameters to produce a power spectrum similar to recent best
fits to observations \citep{2006JCAP...10..014S}:  $\sigma_8=0.85$, $n=0.96$, 
and $\Gamma=0.15$.  I reiterate that this $\Gamma$ is chosen to give the 
closest possible match to the best fit power spectrum from 
\cite{2006JCAP...10..014S}, i.e., it has nothing to do with the Einstein-de 
Sitter background evolution, and is less than $\Omega_m h$ in the concordance 
model because of
broad-band baryonic power suppression which is not otherwise included in the
transfer function of \cite{1984ApJ...285L..45B}. 
\begin{figure}
\resizebox{\textwidth}{!}{\includegraphics{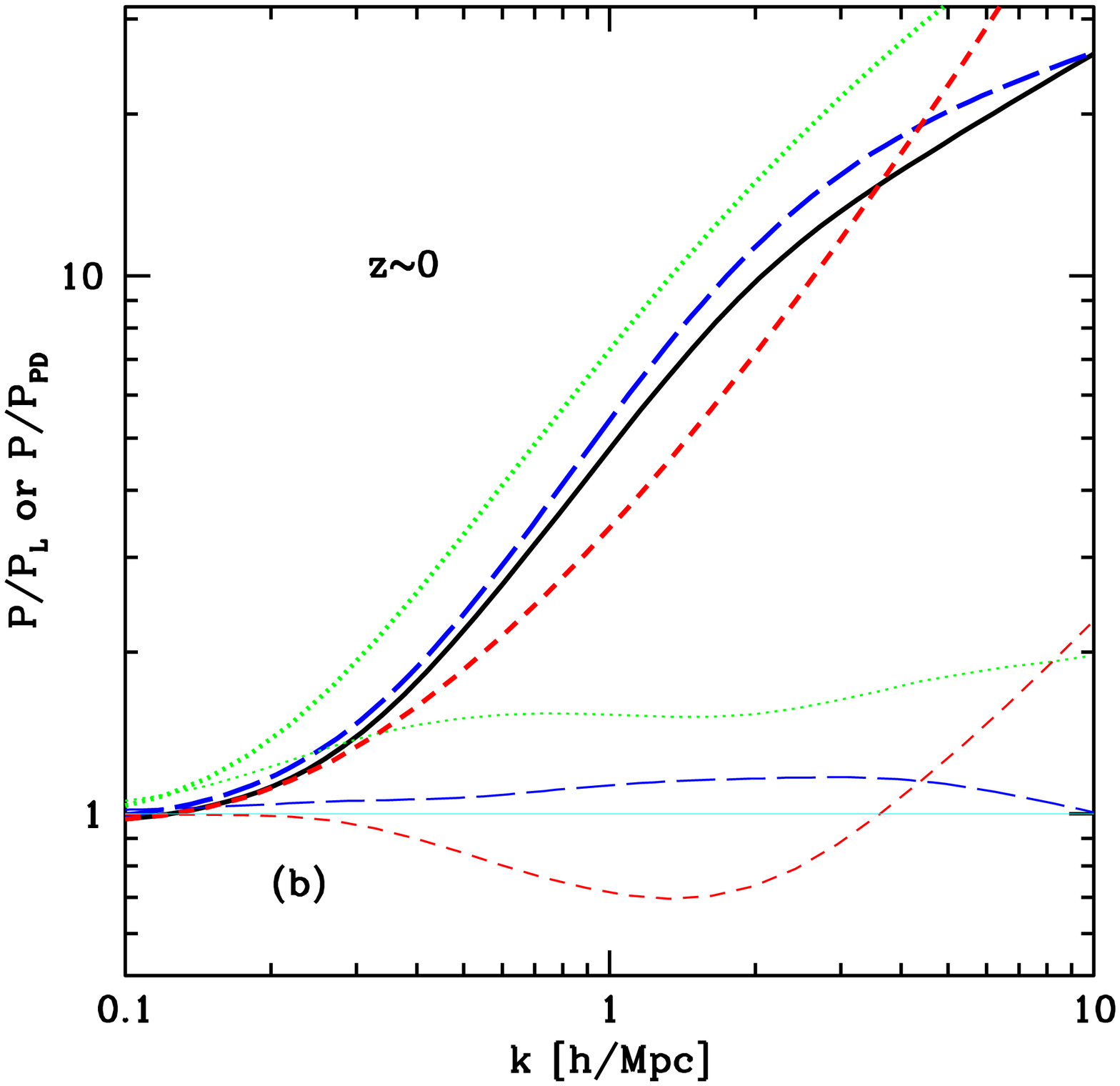}}
\end{figure}

Figure \ref{z0}(a) shows essentially perfect (better than percent level) 
agreement at $z=0$ between RGPT 
and the Peacock and Dodds fitting formula 
\cite{1996MNRAS.280L..19P} for 
$k\lesssim 0.3\ihmpc$.  This does {\it not} mean that RGPT is perfect --
it is likely that the agreement with 
PD96 is partially a coincidence, because we have no 
reason to expect their formula itself to be accurate 
to this level.  In fact, because of the approximation discussed above 
related to decaying modes, RGPT must inevitably have some error as well.
I note here that \cite{2003MNRAS.341.1311S}'s HALOFIT
formula is not suitable for testing perturbation theory at low-$k$ 
because the fit was actually constrained by PT calculations for 
$k<0.1\ihmpc$ (R. E. Smith, private 
communication).   
The fitting formula is a smooth function, so it is not 
surprising that it does not agree perfectly with standard PT at 
$k<0.1\ihmpc$, i.e., if the PT power is generally greater than the simulation  
power, the region near $k=0.1\ihmpc$ will be some average of the two, because a
step function is not allowed.  It is not surprising that this imperfect fit 
would be missed by \cite{2003MNRAS.341.1311S}, because the deviation is
$<5$\%, i.e., smaller than their claimed accuracy.
Additionally, the $k<0.1 \ihmpc$ power was
not constrained exclusively by PT, because the criterion for using scale free 
simulation results was different (R. E. Smith, private
communication).  In any case it is clear that RGPT is an
improvement over standard PT.

Figure \ref{z0}(b) shows the comparison at $z=0$ again, with rescaled axes, 
this 
time focusing on higher $k$, including deep in the non-linear regime.
I note optimistically that, while the prediction is not precise at high $k$, it
continues to be a dramatic improvement over linear theory, i.e., even when
the non-linear power is a factor of $\sim 10$ greater than the linear power,
the perturbation theory result correctly accounts for most of the difference
(admittedly, this was also true of standard PT).
If similar improvements are obtained by removing approximations or simply
with each additional order of perturbation 
theory, one could imagine converging quickly to a precise result.
However, it is possible, or even likely, that missing higher moments of the 
velocity distribution will undermine the calculation at high $k$.  

One of the notable features of the results is their convergence at 
high $k$ to the $d\ln P/d\ln k\simeq-1.4$ fixed point of Eq.~(\ref{eqbeta}),
where the 2nd order PT term 
vanishes for a pure power law power spectrum \citep{1996ApJ...473..620S}. 
At $z=0$ this limit is reached for $k\gtrsim 0.3\ihmpc$, while at earlier 
times it is only reached at higher $k$ (e.g., $k\gtrsim 3\ihmpc$ at $z=3$). 
Qualitatively, this seems like a positive development.
Whatever small-scale structure exists initially, it appears to be effectively 
taken out of play.  This may alleviate one common worry about this kind of
PT, that one is integrating over high-$k$ modes that are highly non-linear, 
violating the premise of the perturbation theory.  
This fixed-point behavior provides another reason to hope that higher order
calculations can significantly extend the range of accuracy of the calculation:
it may be that the high $k$ behavior of $P(k)$ can be thought of as a slow 
rolling with scale, as higher order terms become important,
of this kind of fixed-point power law.  The fixed point will also be modified
by the full inclusion of decaying modes.   

Figure \ref{z1} shows the comparison at $z=1$, a typical redshift for planned
galaxy redshift surveys aimed at measuring the baryonic acoustic oscillation 
feature \citep{2005astro.ph..7457G}.  
As at $z=0$, RGPT agrees well
with the fitting formulas in the mildly non-linear regime, although now the
agreement is better with HALOFIT than PD96 at the high 
$k$ end of this regime.
An optimistic interpretation of this figure would be that PD96
is more accurate at low $k$, because it was not
constrained by standard PT, but HALOFIT becomes more accurate at higher $k$,
as the authors \cite{2003MNRAS.341.1311S} claim it should be.  Dedicated 
numerical 
simulations and/or higher order PT calculations testing convergence are needed
to determine the true power. 
\begin{figure}
\resizebox{\textwidth}{!}{\includegraphics{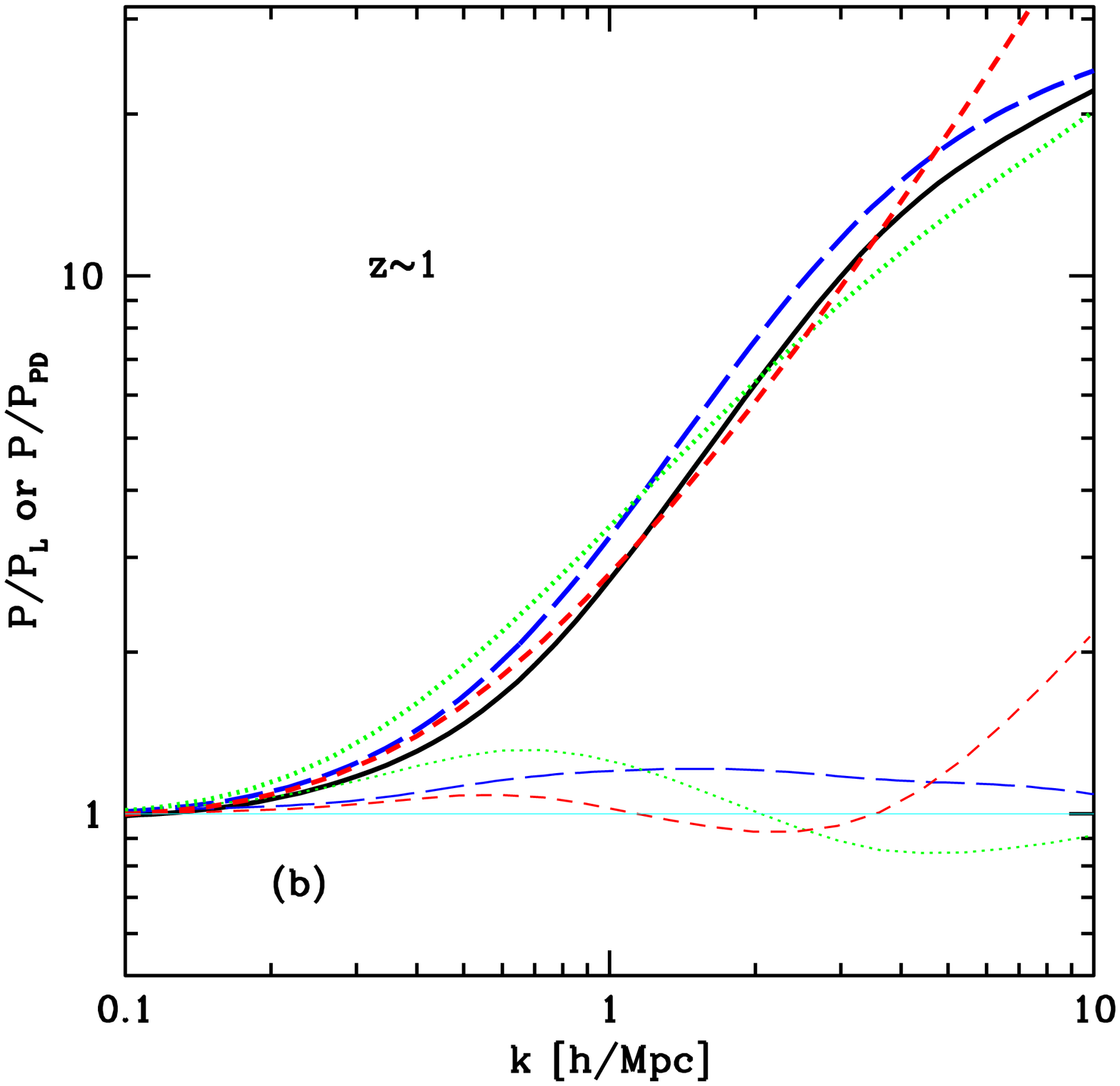}}
\end{figure}
Figure \ref{z1}b again shows that the RGPT prediction traces the fitting 
formulas reasonably well deep into the highly non-linear regime.  In this case,
the agreement is actually as good as the agreement between the two fitting
formulas, out to $k\sim 5\ihmpc$.

\begin{figure}
\resizebox{\textwidth}{!}{\includegraphics{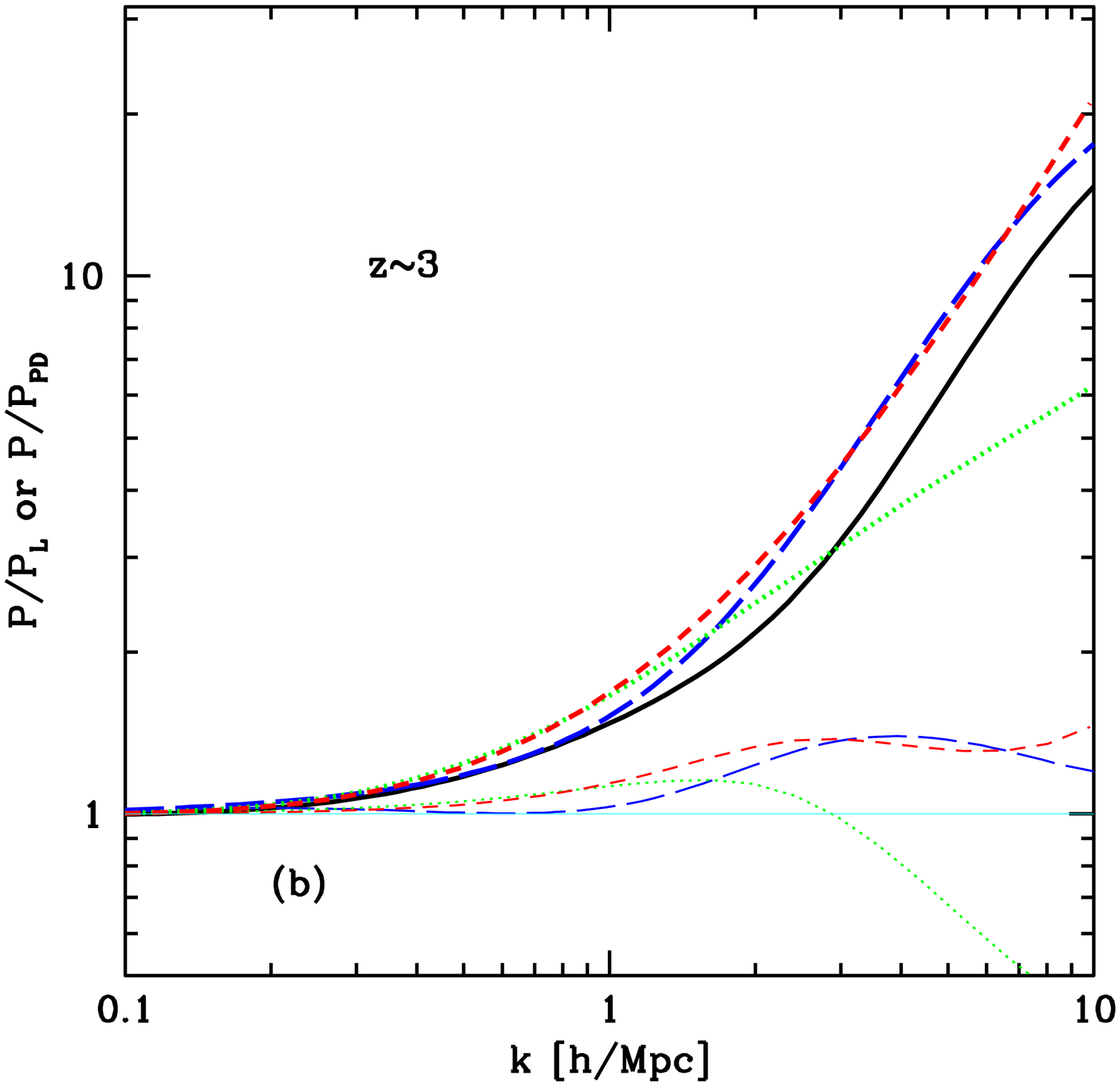}}
\end{figure}
Moving on to $z=3$, relevant to the \lyaf\ 
\citep{1999ApJ...518...24M,2000ApJ...543....1M,
2006ApJS..163...80M}, 
and again potentially future 
galaxy redshift surveys \citep{2005astro.ph..7457G}, we see clear 
deviation between the PT predictions and the fitting formulas, even in the 
weakly non-linear regime.  This is especially puzzling considering that the RG 
and standard version of PT actually give very similar predictions in the weakly
non-linear regime.  
This agreement means that the higher order terms that are
being effectively re-summed by the RG calculation are not important.
Considering the substantial disagreement between the two fitting formulas at
very low $k$, I am not ready to conclude that PT is wrong.  It may be that 
the fitting formulas were not well-calibrated in this regime.  Focused 
simulations are needed to test this. 
Deeper in the non-linear regime at $z=3$, Fig. \ref{z3}(b) shows that the
effect of renormalization becomes substantial, and that RGPT
agrees quite well with HALOFIT, while both disagree significantly with PD96.
The optimistic reading is that RGPT is very 
successful, but considering my willingness to dismiss HALOFIT in the weakly 
non-linear regime, it is probably best to wait for more simulations and/or
higher order PT calculations before becoming too pleased with RGPT.

\begin{figure}
\resizebox{\textwidth}{!}{\includegraphics{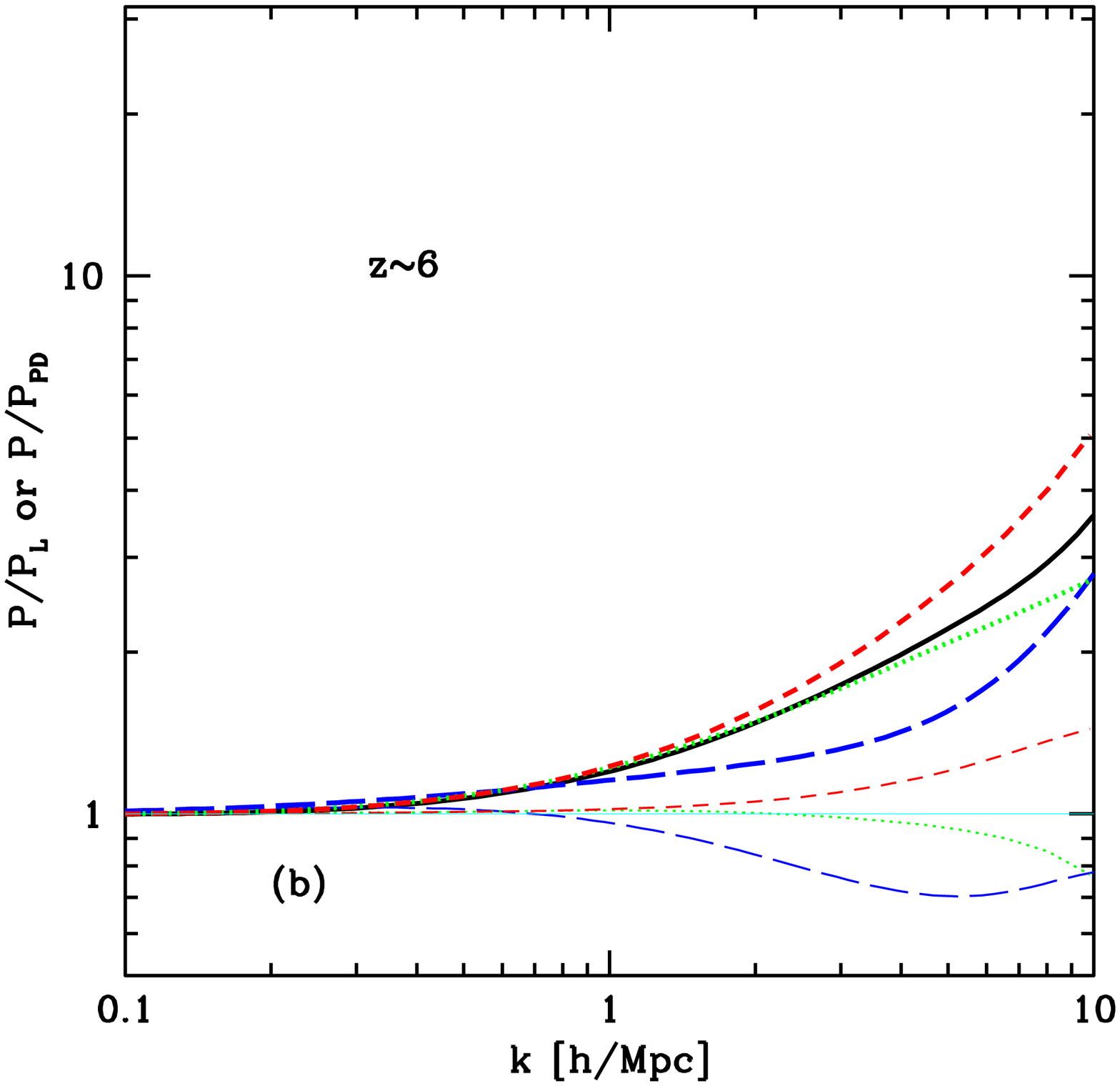}}
\end{figure}
Finally, we consider $z=6$.  In the weakly non-linear regime shown in Fig.
\ref{z6}(a), RG and standard PT are essentially identical, and agree quite
well with PD96.  HALOFIT differs substantially, so 
again
the results are ambiguous.  The same can be said for the non-linear
regime in Fig. \ref{z6}(b), where the fitting formulas disagree substantially.

\section{Discussion and Speculation \label{secdiscuss}}

The central result of this paper is Eq.~(\ref{eqbeta}), the RG equation for
the renormalized power spectrum, demonstrated in Fig. \ref{z0}(a) to give 
good results for the power spectrum in the quasi-linear regime at 
low redshift, where standard PT performs poorly.
It is clear from Figs. \ref{z0}-\ref{z6} that the RG improvement to standard 
PT is generally helpful, in both the weakly and strongly non-linear 
regimes.  Except in the most challenging case of $z=0$ and high $k$,
the simulation-calibrated fitting formula predictions of Peacock and Dodds
\cite{1996MNRAS.280L..19P} and HALOFIT \cite{2003MNRAS.341.1311S} are 
sufficiently 
contradictory that it is not completely obvious that they are more reliable 
than RGPT. 

It is possible that this method will turn out to be a computational curiosity,
maybe leading to improved a posteriori understanding of what we see in 
simulations, but unable to achieve sufficient accuracy for practical 
problems, and thus leading 
to little fundamental change in how we carry out precision cosmology 
measurements. (It may appear that the method is guaranteed to be useful for 
describing weakly non-linear galaxy clustering, but this could still be foiled 
by bias and redshift-space distortions, which I discuss below.)  On the other 
hand, 
I hope that this line of work is only in its infancy, and will lead to
a diversification of computational methods, beyond standard N-body simulations. 
I am distinguishing here between systematic approximation methods and 
heuristic ones like the halo model, which rely heavily on calibration by 
simulations. 
One of the implicit goals of this work has been to avoid any 
approximations where the path to improved accuracy is unclear.
A good quality for an approximation to have is a
clear limit in which it can be expected to return to the exact result.
Note that standard simulations are themselves exactly the kind of systematic 
approximation that I would like to see more of.  Simulating a finite volume  
with a finite number of particles and limited force resolution are all 
approximations, but each can be tested for convergence as their controlling
parameters are changed.

With the hope of stimulating further work, I now discuss a variety of possible
directions that could be pursued, although I can't guarantee that all of these
are good ideas:

The first priority should be to include the 
velocity-velocity and velocity-density power spectra as independent functions
to be renormalized, as I discussed extensively in \S \ref{secexplanation}. 
The main requirement for doing this properly is a straightforward 
generalization of the standard PT results to allow for 
general initial conditions, including decaying modes.  
The $n\simeq-1.4$ fixed-point power spectrum will be modified, hopefully to a
more interesting, accurate fixed-point shape.

After that, brute force computation to higher order in perturbation 
theory should produce more accurate results.  The 
fact that the current calculation is a substantial improvement over linear 
theory at all $k$ (as opposed to diverging wildly at some point), well beyond 
the scale where it is no longer very precise, makes me 
hopeful that each additional order could extend the effective scale 
significantly.  The possibility of estimating the errors in the calculation by
comparing results at different orders is an equally strong motivation for going
to higher order.  [In the same sense, the present calculation should show where
linear theory breaks down much more accurately than the common practice of
guessing based on the value of $\Delta^2(k)=k^3 P_L(k)/2 \pi^2$.]

These calculations give clear motivation and targets for high accuracy 
simulation work.   RGPT, being a first principles method 
containing the possibility of internal error control,
may in places lead the comparison rather than simply being
tested itself.  If a higher order of PT is computed, and the results agree 
between different orders to
some level of precision, it is reasonable to expect that the result is correct,
independent of simulations.  (Missing higher moments of the particle velocity
distribution function may be a loophole in this argument, as I discuss below.)
With  the fixed-point concept in mind, it may be interesting to
look more carefully at the evolution of the logarithmic slope of the
power spectrum with time and scale in simulations.

Abandoning the first principles philosophy, RGPT results may provide a useful
template for a new simulation fitting formula.  Because the errors in RGPT are
not too large, and should be slowly varying functions of $k$ and cosmological 
parameters, the new formula could be just a very simple parameterization of 
corrections to RGPT computed using simulations.

One can obviously apply the techniques of this paper to higher order 
statistics such as the bispectrum 
\citep{2005MNRAS.361..824G,2006A&A...456..421B,2006PhRvD..74b3522S}, 
trispectrum,  etc. \citep{2006ApJ...649...48R}.  As discussed 
in \S \ref{secexplanation}, one can 
consider promoting the leading order for these statistics to the more generally 
possible $\mathcal{O}(\delta_1^3)$ and $\mathcal{O}(\delta_1^4)$ (for the 
bispectrum and trispectrum, respectively).  
Bispectrum results can be compared to the simulation
fitting formula of \cite{2001MNRAS.325.1312S}.
If the trispectrum computation is sufficiently accurate, RGPT should be useful 
for computing expected statistical errors on the
power spectrum, which is notoriously painful to do with simulations 
\citep{2006MNRAS.371.1188H}.  Similarly, errors on higher order statistics
may be computable.

Mostly for convenience, I have assumed that the only effect of deviations from 
Einstein-de Sitter background evolution is to change the linear growth factor.
This is probably sufficient in the weakly non-linear regime, but if RGPT is to 
achieve much accuracy at higher $k$, this assumption inevitably must be
relaxed.  This is clear, even though the corrections in standard PT are very 
small, because the dependence of the power spectrum on the equation of state
of dark energy, $w$, at fixed observation-time linear theory power 
\citep{2006MNRAS.366..547M}, can not be reproduced under the approximation used
here.  There is simply no way the dependence quantified in 
\cite{2006MNRAS.366..547M} can enter the current calculation.  Fortunately,
even small changes in the right hand side of Eq.~(\ref{eqbeta}) can 
lead to significant changes in the solution of this non-linear equation.
It will also be necessary to move beyond the very simple time dependence in
Eq.~(\ref{Pspeceq}) if one wants to include any non-gravitational effects 
that break this form (e.g., gas pressure).

To describe direct tracers of gas, like the \lyaf\ or SZ, gas dynamics could be 
included in the calculation.  Even for weak lensing, which traces mass
directly, gas
physics has been shown to matter at a level problematic for high precision
experiments \citep{2006ApJ...640L.119J}.  Very rudimentary pressure 
approximations should be relatively straightforward to include 
\citep{1998MNRAS.296...44G}.  While it may be that very sophisticated 
implementations of heating and cooling are hopeless,
one should remember that there is some non-trivial 
inclusion of non-linearity in this calculation, so this is at least worth
considering.

To describe clustering of galaxies, or other unavoidably inexact tracers of 
density, some concept of bias will need to be included in the calculation.  
I think this may be the area where RGPT is most likely to provide a real 
fundamental improvement
over simulations, or at least complement to them.  With a large but probably 
achievable amount of computer power, one can simulate the power 
spectrum of dark matter and its dependence on a relatively small number of 
cosmological parameters more or less perfectly.  This will not be possible
in the foreseeable future for galaxies, so robust, high precision cosmological 
measurements using
them will rely on developing a comprehensive understanding of how relatively
microscopic galaxy formation details translate into large-scale clustering
patterns.  
While the separation of scales probably is not clear enough to make
the analogy perfect, this is reminiscent of the classic uses of the
renormalization group to understand how complex microscopic theories lead to 
simple macroscopic behavior \citep{1983RvMP...55..583W}.   
The simplest way to introduce bias is to follow 
\cite{1993ApJ...413..447F,1998MNRAS.301..797H} in writing the galaxy density
fluctuation, $\delta_g$, as a Taylor series in the mass density,
$\delta_g=\sum_i b_i \delta^i /i!$, which leads to an expression for the PT
power spectrum of galaxies involving a few more terms than the mass power
spectrum,  but not fundamentally more 
complicated.  This path has been followed by \cite{2006PhRvD..74j3512M}.
Much more ambitiously, one could write local formation rate equations for a
galaxy density field and apply perturbation theory to them 
(something like this was discussed using linear theory in 
\citep{1998ApJ...500L..79T}).  
For example, $\dot{\rho}_g=\rho~f(\rho)$, with 
$\dot{\rho}=-\dot{\rho}_g$ to represent 
depletion of gas, and with both components obeying the usual 
gravitational evolution equations.  The mean galaxy density as a
function of time  would be an output of the calculation as well as 
fluctuations.  The goal here would not be to write down a realistic local model
and compute exact predictions as much as to see what kind of relations between
observables can be generically expected.
An intriguing possibility is that one could show that, regardless of the 
small-scale galaxy formation details, bias can only take certain universal 
forms, described by a small number of free parameters.  Note that the validity
of the scale-independent bias assumption has never really been proven even in 
the perfectly linear regime (see \cite{1998ApJ...504..607S}
for one attempt).  This entire discussion of bias of galaxies applies equally
well to many other traces of density, e.g., the \lyaf\
\citep{2003ApJ...585...34M}.

For many applications, redshift space distortions will need to be included.
Like bias, these should be relatively straightforward to include in the present 
RG formulation using the extension of \cite{1987MNRAS.227....1K}'s linear 
theory calculation described by \cite{1998MNRAS.301..797H}.  
\cite{2004PhRvD..70h3007S} discusses the imperfections of the 
\cite{1987MNRAS.227....1K} approach, and proposes alternative ideas which may 
be useful.

It may be interesting, as a computational curiosity, to see if RGPT can give 
sensible results for $n\geq -1$ power law initial conditions.  In this case the 
integrals giving the first correction in standard PT truly diverge, in the
sense of being infinite rather than just large \citep{1996ApJ...473..620S}.
The integrals can be cut off at wavenumber $k_c$, but the results then 
depend on $k_c$, which is arbitrary.
One may be able to show that, using the RG as described in this paper, the 
results will converge as $k_c$ is taken to infinity.  The small-$A_\star$
integration of Eq.~(\ref{eqbeta}) will need to be done increasingly 
carefully as $k_c$ is increased, but hopefully convergence to the $n=-1.4$
fixed point will quickly erase all trace of the cutoff.
Note that $k_c$ could affect the normalization of the asymptotic power law, 
which would spoil this idea (e.g., it may only work over some limited
range of $n$). 

One potential limitation in RGPT as presented here is the absence of higher
moments of the velocity distribution function, e.g., the dispersion in the 
velocities of particles around their mean velocity at a point.
I wrote the Vlasov equation (\ref{eqvlasov}) instead of cutting straight to 
the hydrodynamic equations to point out this problem, shared by standard PT  
(subsequently discussed by \cite{2006astro.ph.10336A}). 
Dropping higher moments of the velocity distribution,
also known as the single-stream approximation, is not really an added
approximation
in standard Eulerian perturbation theory.  
This is surely known to experts (e.g., \cite{1998A&A...335..395B}), but may not
be universally recognized.  
One can easily include these moments as variables in the calculation, for 
example multiplying the Vlasov equation by $p_i p_j$ and integrating over 
$p$ gives (after some substitution to remove bulk velocity terms)
\begin{equation}
\frac{\partial \sigma^{i j}}{\partial \tau}+2~ {\mathcal H}~\sigma^{i j}+
v^k~ \partial_k \sigma^{i j}+\sigma^{j k}~ \partial_k v^i+
\sigma^{i k}~ \partial_k v^j+
\frac{\partial_k\left[\left(1+\delta\right) q^{i j k}\right]}{1+\delta}=0 ~,
\end{equation}
where $\partial_k=\partial/\partial x^k$, 
$\sigma^{i j}(\vx,\tau)\equiv\left<\delta v^i~\delta v^j\right>$,
$q^{i j k}(\vx,\tau)\equiv\left<\delta v^i~\delta v^j~\delta v^k\right>$,
with $\delta v^i$ the deviation of a particle's velocity from the local 
mean velocity (note that a $\sigma^{i j}$ term would appear in 
Eq. \ref{eqeuler} if I had not dropped it).
Unfortunately, the 
lowest order evolution equations contain
only the Hubble drag term, so any initial velocity dispersion will 
self-consistently disappear from the perturbation theory.
Furthermore, higher order equations contain no source terms, meaning that if 
the velocity dispersion is initially zero it can never become non-zero.  
In fact, \cite{2001A&A...379....8V} showed that even perturbation theory using 
the distribution function directly leads to the same result.
This is a vexing problem
because stream crossing is clearly ubiquitous in the real Universe; however,
\cite{2001NYASA.927...13S} 
estimated that it only becomes significant at $k\gtrsim 1 \ihmpc$, so
it may not be fatal for intermediate scales.
It seems likely that, if these
variables really matter, there should be some way to recover them using a RG
method.  The basic idea is that, no matter how small the effect is in the bare
perturbation theory, if it is dynamically relevant it will grow through a RG 
equation like Eq.~(\ref{eqbeta}).  
Note that even WIMP cold dark matter has a
very small seed velocity dispersion \citep{2004MNRAS.353L..23G}.
The solution is probably to simply retain the dispersion variable in the 
calculation, even though 
it is decaying, in the same way I have argued above that one should retain
the decaying combination of $\delta$ and $\theta$ -- because it can be rapidly 
fed by higher order terms.
Vorticity, $\vnabla \times \vv$, falls out of
the standard PT calculation for the same reason velocity dispersion does
\citep{2002PhR...367....1B}.  
The standard calculation makes use only of the velocity 
divergence, $\vnabla \cdot \vv$.  A method that successfully reintroduces 
velocity dispersion may also work for vorticity. 

I started this line of work with the intention of using a significantly 
different type of renormalization group method, based on integrating out 
(averaging over) small-scale (high-$k$) modes while modifying the equations 
describing 
evolution of the large-scale modes in a way that preserves their statistics
\citep{1994RvMP...66..129S}.
This method usually starts with a path integral formulation of the problem.
\cite{2001A&A...379....8V} presented a path integral approach to large-scale
structure using the full distribution function, and it should not be too hard
to write something similar for the usual moments of the distribution function.
Velocity dispersion should arise inevitably in this approach,  
because smoothing a dispersionless
field obviously produces dispersion.  It is less inevitable that 
this dispersion will actually fundamentally change the outcome of the 
calculation, because the effect 
of these velocities is already present in the standard PT calculation.   
Among other possibly interesting features, the noise discussed in previous RG 
approaches to large scale structure
\citep{1997EL.....38..637B,1999A&A...344...27D} may be generated naturally
using this approach (i.e., if high-$k$ modes are eliminated, the time 
evolution of the remaining modes will no longer be perfectly deterministic).  

Further work could elucidate the connection between this work and the
re-summation method of \cite{2006PhRvD..73f3520C,2006PhRvD..73f3519C}.  
Related renormalization group procedures, starting from general
relativistic perturbation theory, are used by 
\cite{2000PhRvD..62j4010N} and \cite{2005astro.ph..6534K} to compute the 
back-reaction of inhomogeneities on the global evolution of the Universe.

Finally, the success of the RGPT calculation in this paper, which leads to
what looks like a time evolution equation for the power spectrum, suggests
that it might be useful to consider the exact 
time evolution equation for the power spectrum, 
$\dot{P}(k)\propto 2~{\rm Re}\left<\dot{\delta}_\vk~\delta_{-\vk}\right>$, 
where $\dot{\delta}_\vk$ is obtained from the usual evolution equations.
The usual argument against this would be that 
$\left<\dot{\delta}_\vk~\delta_{-\vk}\right>$ involves higher order terms like
the bispectrum, which then require their own time evolution equations, and so
on, leading to an infinite hierarchy of equations. 
Firstly, this could be a useful way to look at perturbation theory, 
especially if
we are going to be renormalizing statistics like the power spectrum instead of 
the field itself.  Second, one may be able to make progress by simply 
solving these equations numerically, in the same sense that simulations
are usually used.
One could, for example, truncate the hierarchy by replacing the 
connected spectra at some level by the appropriate functions of lower order 
spectra given by perturbation theory, and look for convergence in
the results as higher order terms are added.
The advantage of working with statistics of the density field rather than the
density field itself is enormous.  One 
obtains the desired results directly rather than averaging over fluctuations
in simulations, where there is usually an unavoidable conflict between
the need for large boxes to limit finite volume effects like sample variance, 
and the need
for high force and mass resolution to properly compute small-scale structure.
The fact that we are working with continuous, relatively slowing varying 
functions, rather than wildly fluctuating fields, allows many fewer points to
be computed, e.g., we can compute a power spectrum over many decades of 
dynamical range by interpolating between a few tens of logarithmically spaced
computation points when millions or billions of particles would be 
needed to cover the same range with a simulation.
Tasks that at first glance seem arduous, e.g., parameterizing the trispectrum 
or even higher spectra and setting up their evolution equations, should be 
considered in contrast to the decades of person-power expended on simulations. 

\acknowledgements

I thank Lev Kofman, Niayesh Afshordi, and Roman Scoccimarro for helpful 
discussions, and Robert Smith and Ue-Li Pen for comments on the manuscript.
Some computations were performed on CITA's
McKenzie cluster which was funded by the Canada Foundation for Innovation and
the Ontario Innovation Trust \citep{2003astro.ph..5109D}.

\bibliography{cosmo,cosmo_preprints}

\end{document}